\documentclass[a4paper]{book}
\usepackage[pdftex]{graphicx}
\usepackage{nano2cmr}
\begin{document}

\pnum{}
\ttitle{Linear and nonlinear coupling of quantum dots in microcavities}
\tauthor{{\em F.~P.~Laussy}, E.~del Valle, A. Gonzalez-Tudela, E. Cancellieri, D. Sanvitto and C. Tejedor.}


\ptitle{Linear and nonlinear coupling of quantum dots in microcavities}
\pauthor{{\em F.~P.~Laussy},$^1$ E.~del Valle,$^1$ A. Gonzalez-Tudela,$^2$ E. Cancellieri,$^2$ D. Sanvitto$^2$ and C. Tejedor.$^2$}

\affil{$^1$ School of Physics and Astronomy, University of Southampton, SO17{$\,$}1BJ, Southampton, United Kingdom,\\
$^2$ Universidad Aut\'onoma de Madrid, 28$\,$049, Madrid, Spain.
}

\begin{abstract} { We discuss the topical and fundamental problem of
    strong-coupling between a quantum dot an the single mode of a
    microcavity. We report seminal quantitative descriptions of
    experimental data, both in the linear and in the nonlinear
    regimes, based on a theoretical model that includes pumping and
    quantum statistics.}
\end{abstract}

\begindc 

\index{First author}
\index{Principal}   
\index{Next author}
\index{Last author}

\section*{Introduction}

\begin{figure}
  \centering
  \includegraphics[width=.7\linewidth]{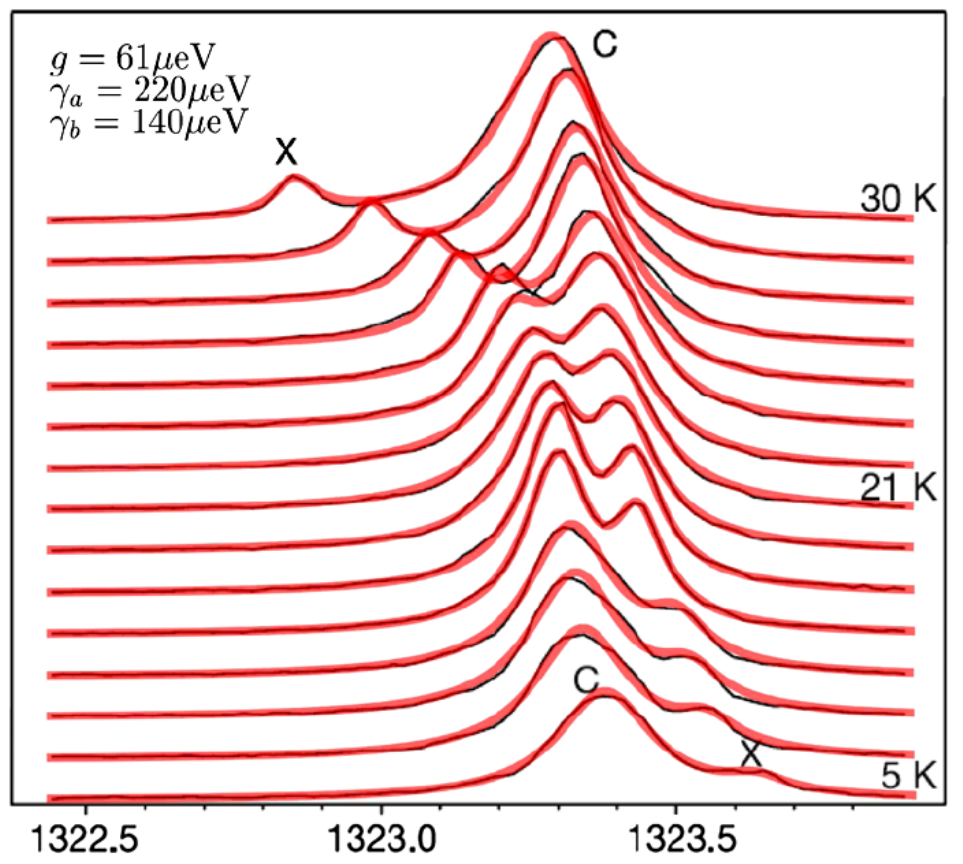}
  \caption{Fit in the linear regime: data of Reithmaier \emph{et
      al.}~\cite{reithmaier04a} with---superimposed in red---our
    global fit~\cite{laussy08a}.}
  \label{fig:SunFeb14210147GMT2010}
\end{figure}

After its pioneering observation in micropillars~\cite{reithmaier04a}
and in photonic crystals~\cite{yoshie04a} in 2004, strong-coupling of
light and matter is now commonplace in zero-dimensional
nanostructures~\cite{hennessy07a,press07a,
  nomura08a,kistner08a,laucht09b,dousse09a}. This physics is
primordial both for fundamental research and technological
applications, from thresholdless lasers and new light-sources to
quantum information processing. We have given the first quantitative
description of experimental data~\cite{laussy08a} (see
Fig.~\ref{fig:SunFeb14210147GMT2010}), by providing a general model of
quantum modes coupling~\cite{laussy09a}. Our model extends the
atomistic description, that limits to the excited state of the atom in
an empty cavity as an initial condition. In semiconductor
microcavities, however, photoluminescence measurements are typically
made in the steady state established by a continuous, incoherent
pumping. We take these specificities into account and: i) recover the
spontaneous emission case of an arbitrary initial state in the limit
of vanishing pumping, thereby providing a complete, self-consistent
and general description of light-matter coupling, and ii) include
dynamical effects of quantum statistics for non-vanishing pumping,
such as Bose stimulation and Pauli blocking.

\section {Linear regime}

When pumping is very small, so that the system is most of the time in
vacuum, and occasionally excited, its photoluminescence spectrum is
that of spontaneous emission of the state that results from the
averaged excitation. In a microcavity, excitation is typically sought
to be of the quantum dot itself, by mean of electron-hole pairs
relaxation from off-resonant pumping~\cite{averkiev09a}. As such, its
photoluminescence spectrum would recover that of the atomic
literature~\cite{carmichael89a}, up to a technical correction that
consists in computing the cavity mode spectrum (for a microcavity)
rather than the direct atomic de-excitation (for an optical
cavity). However, various mechanisms result in an effective
microcavity pumping, where photons are directly injected into the
coupled light-matter system. One vivid scenario is that the quantum
dot of interest, in strong-coupling with the cavity, is a ``lucky''
one---well positioned in the optical field, with adequate coupling
strength, etc.---but is surrounded by many other dots, less
efficiently coupled to the cavity (in weak-coupling). These are also
excited by electron-pairs ideally intended for the strongly-coupled
dot only. The other dots can release efficiently (by Purcell
enhancement) and with no correlations (in a Markovian approximation)
their excitation in the cavity mode and as such bath the light-matter
system of interest in a photonic environment. As the initial state
(and effective quantum steady state) affects dramatically the spectral
shapes, it is important to take into account both excitation channels
to reproduce experimental data, in particular as effects of pumping
are known to produce nontrivial phenomenology even at a qualitative
level~\cite{keldysh06a}. Most strikingly, at resonance, neither a
doublet is a guarantee of strong-coupling nor a singlet is an evidence
of weak-coupling. The analysis should be done at the level of spectral
shapes rather than for the anticrossing of the maxima, for which there
is no closed expression~\cite{gonzaleztudela10a}.

\section {Nonlinear regime}

When pumping is non-negligible, quantum statistics is to be added to
the previous considerations. In the case of a large or elongated
quantum-dot, where the electron and hole can bind as an exciton, the
underlying statistics is that of Bose-Einstein~\cite{laussy06b}. This
results in stimulated emission and line narrowing with increasing
pumping (Schallow-Townes effect). If, on the other hand, the electron
and hole are quantized separately in a small quantum dot, Pauli
exclusion prevents other electrons and holes to populate the already
excited dot, and the statistics of Fermi-Dirac rules the dynamics. As
a conclusion, increasing excitation can lead to a variety of rich
nonlinear effects, in the wake of Jaynes-Cummings physics (strong
coupling of a boson and fermion). An important expected manifestation
is full-field quantization, that results in a series of peaks at
anharmonic frequencies $\pm(\sqrt{n}\pm\sqrt{n-1})$ when~$n$ excitations
are in the system~\cite{laussy09b}. State of the art technology does
not yet allow to resolve clearly this fine-structure in
semiconductors, although we find that a careful analysis in a
situation with significant cavity pumping, could evidence
manifestations of nonlinearities at the quantum
level~\cite{delvalle09a}.

\begin{figure}
  \centering
  \includegraphics[width=.6\linewidth]{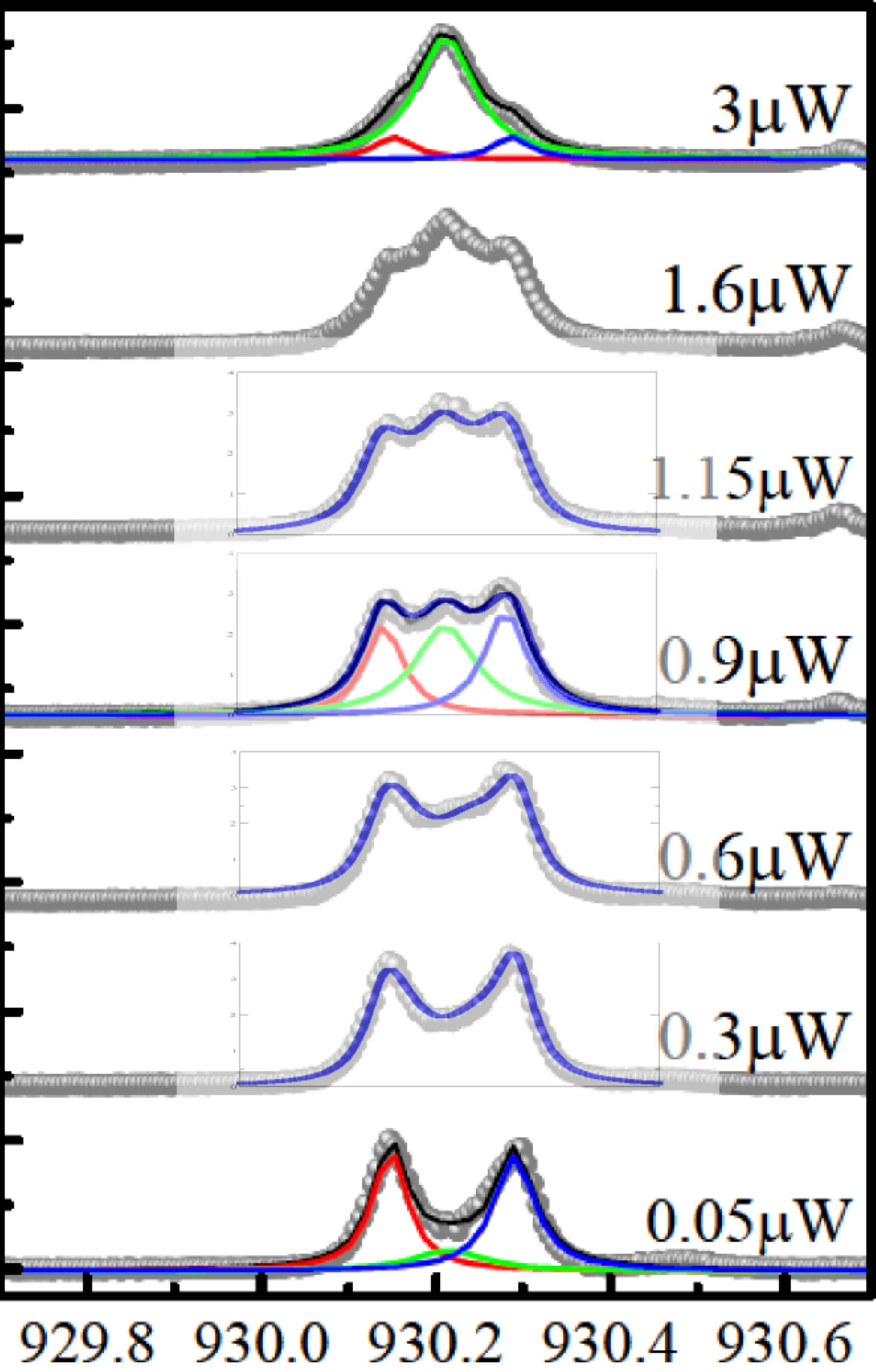}
  \includegraphics[width=.7\linewidth]{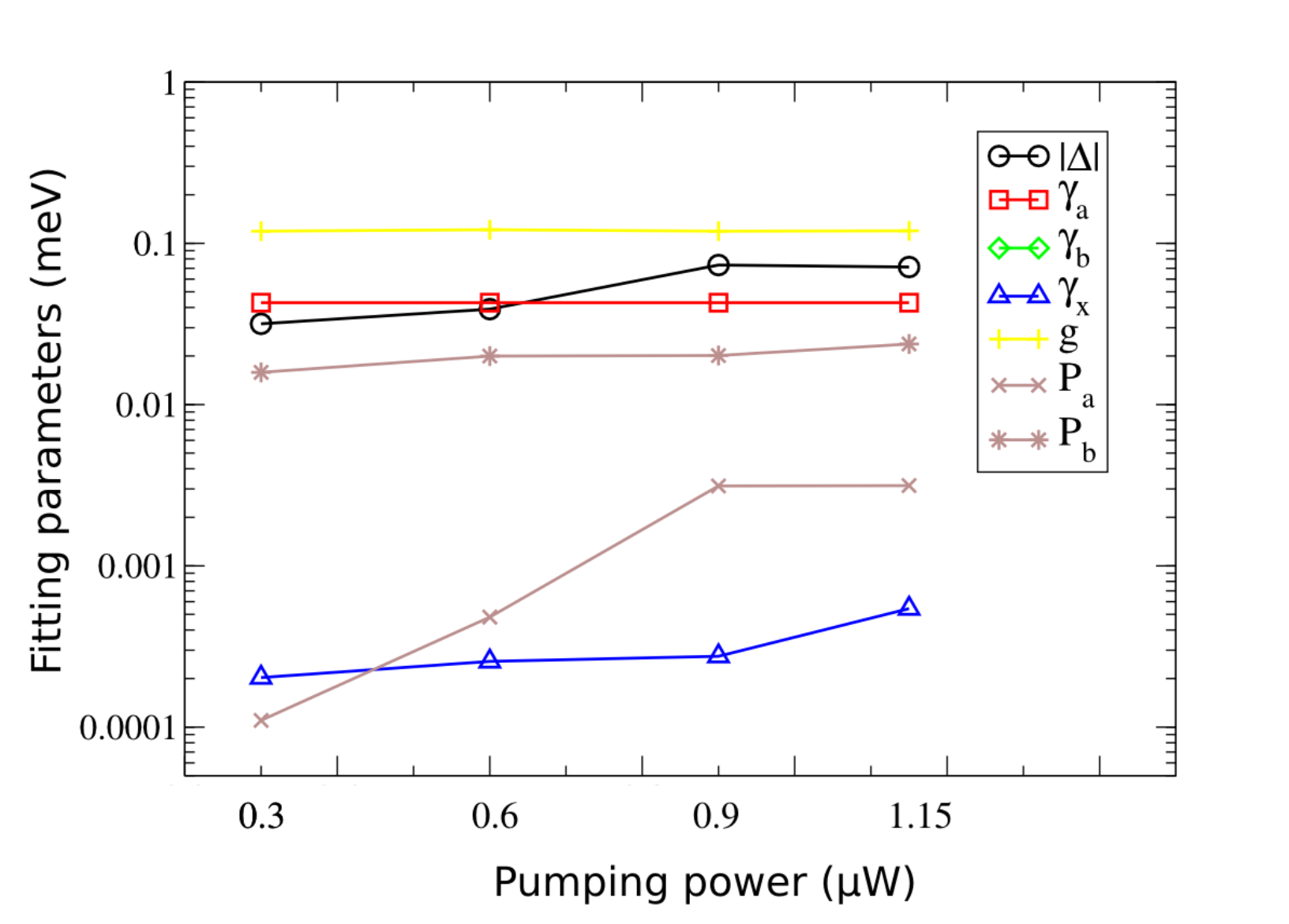}
  \caption{Fit in the nonlinear regime: data of Ota \emph{et
      al.}~\cite{ota09b} with---superimposed in blue for the four
    central panels---our global fit (fitting parameters appear below)
    with a fermion model~\cite{delvalle09a} including
    dephasing~\cite{gonzaleztudela10b}.}
  \label{fig:SunFeb14214113GMT2010}
\end{figure}

Another important expected manifestation of a two-level system brought
in the nonlinear regime by high pumping is lasing. Recently, the
transition from vacuum strong-coupling to lasing has indeed been
reported~\cite{nomura10a}. In another, related
work~\cite{ota09b}, this transition was observed to pass through a
stage where a triplet is formed by appearance of a peak at the cavity
mode frequency that subsequently overtakes the polariton modes as the
system enters into lasing (see
Fig.~\ref{fig:SunFeb14214113GMT2010}). Various explanations have been
advanced for spectral triplets of this
type~\cite{hennessy07a,ota09b}. We offer one that, including a term of
dephasing---which has been demonstrated to play a key role with
increasing pumping~\cite{laucht09b}---explains the appearance of the
triplet as a melting of the inner transitions between rungs of the
Jaynes-Cummings ladder. Such a transition, if confirmed, would provide
a striking crossover from the quantum realm, where single quanta rule
the dynamics of the system~\cite{khitrova06a}, to the classical world,
where a continuous field (the lasing mode) takes over a small number
of quantum correlators. We have confronted our theory with the
experimental data, again by fitting, but the situation in the
nonlinear regime is significantly more complicated, owing to the lack
of closed expressions for the spectral lineshapes. We have used
genetic algorithm methods to do a global fitting of the data. We find
our proposition to be consistent with the supposed parameters of this
experiment, beside with a neat contribution due to a drift in detuning
more than to dephasing. Beyond supporting claims of quantum
nonlinearities, our work also provides the first quantitative
description of strong-coupling experimental data but now in the
nonlinear and fermionic regime.

\bibliographystyle{apsrev4-1}
\bibliography{Sci,arXiv,neva}

\begin{thebibliography}{10}%
\makeatletter
\providecommand \@ifxundefined [1]{%
 \ifx #1\undefined \expandafter \@firstoftwo
 \else \expandafter \@secondoftwo
\fi
}%
\providecommand \@ifnum [1]{%
 \ifnum #1\expandafter \@firstoftwo
 \else \expandafter \@secondoftwo
\fi
}%
\providecommand \enquote [1]{``#1''}%
\providecommand \bibnamefont  [1]{#1}%
\providecommand \bibfnamefont [1]{#1}%
\providecommand \citenamefont [1]{#1}%
\providecommand\href[0]{\@sanitize\@href}%
\providecommand\@href[1]{\endgroup\@@startlink{#1}\endgroup\@@href}%
\providecommand\@@href[1]{#1\@@endlink}%
\providecommand \@sanitize [0]{\begingroup\catcode`\&12\catcode`\#12\relax}%
\@ifxundefined \pdfoutput {\@firstoftwo}{%
 \@ifnum{\z@=\pdfoutput}{\@firstoftwo}{\@secondoftwo}%
}{%
 \providecommand\@@startlink[1]{\leavevmode\special{html:<a href="#1">}}%
 \providecommand\@@endlink[0]{\special{html:</a>}}%
}{%
 \providecommand\@@startlink[1]{%
  \leavevmode
  \pdfstartlink
   attr{/Border[0 0 1 ]/H/I/C[0 1 1]}%
   user{/Subtype/Link/A<</Type/Action/S/URI/URI(#1)>>}%
  \relax
 }%
 \providecommand\@@endlink[0]{\pdfendlink}%
}%
\providecommand \url  [0]{\begingroup\@sanitize \@url }%
\providecommand \@url [1]{\endgroup\@href {#1}{\urlprefix}}%
\providecommand \urlprefix [0]{URL }%
\providecommand \Eprint[0]{\href }%
\@ifxundefined \urlstyle {%
  \providecommand \doi [1]{doi:\discretionary{}{}{}#1}%
}{%
  \providecommand \doi [0]{doi:\discretionary{}{}{}\begingroup
  \urlstyle{rm}\Url }%
}%
\providecommand \doibase [0]{http://dx.doi.org/}%
\providecommand \Doi[1]{\href{\doibase#1}}%
\providecommand \bibAnnote [3]{%
  \BibitemShut{#1}%
  \begin{quotation}\noindent
    \textsc{Key:}\ #2\\\textsc{Annotation:}\ #3%
  \end{quotation}%
}%
\providecommand \bibAnnoteFile [2]{%
  \IfFileExists{#2}{\bibAnnote {#1} {#2} {\input{#2}}}{}%
}%
\providecommand \typeout [0]{\immediate \write \m@ne }%
\providecommand \selectlanguage [0]{\@gobble}%
\providecommand \bibinfo [0]{\@secondoftwo}%
\providecommand \bibfield [0]{\@secondoftwo}%
\providecommand \translation [1]{[#1]}%
\providecommand \BibitemOpen[0]{}%
\providecommand \bibitemStop [0]{}%
\providecommand \bibitemNoStop [0]{.\EOS\space}%
\providecommand \EOS [0]{\spacefactor3000\relax}%
\providecommand \BibitemShut [1]{\csname bibitem#1\endcsname}%
\bibitem{reithmaier04a}%
  \BibitemOpen
  \bibfield{author}{%
  \bibinfo {author} {\bibfnamefont{J.~P.}\ \bibnamefont{Reithmaier}}, \bibinfo
  {author} {\bibfnamefont{G.}~\bibnamefont{Sek}}, \bibinfo {author}
  {\bibfnamefont{A.}~\bibnamefont{L\"offler}}, \bibinfo {author}
  {\bibfnamefont{C.}~\bibnamefont{Hofmann}}, \bibinfo {author}
  {\bibfnamefont{S.}~\bibnamefont{Kuhn}}, \bibinfo {author}
  {\bibfnamefont{S.}~\bibnamefont{Reitzenstein}}, \bibinfo {author}
  {\bibfnamefont{L.~V.}\ \bibnamefont{Keldysh}}, \bibinfo {author}
  {\bibfnamefont{V.~D.}\ \bibnamefont{Kulakovskii}}, \bibinfo {author}
  {\bibfnamefont{T.~L.}\ \bibnamefont{Reinecker}},\ and\ \bibinfo {author}
  {\bibfnamefont{A.}~\bibnamefont{Forchel}},\ }%
  \bibfield{journal}{%
  \bibinfo {journal} {Nature}\ }%
  \textbf{\bibinfo {volume} {432}},\ \bibinfo {pages} {197} (\bibinfo {year}
  {2004})%
  \bibAnnoteFile{NoStop}{reithmaier04a}%
\bibitem{yoshie04a}%
  \BibitemOpen
  \bibfield{author}{%
  \bibinfo {author} {\bibfnamefont{T.}~\bibnamefont{Yoshie}}, \bibinfo {author}
  {\bibfnamefont{A.}~\bibnamefont{Scherer}}, \bibinfo {author}
  {\bibfnamefont{J.}~\bibnamefont{Heindrickson}}, \bibinfo {author}
  {\bibfnamefont{G.}~\bibnamefont{Khitrova}}, \bibinfo {author}
  {\bibfnamefont{H.~M.}\ \bibnamefont{Gibbs}}, \bibinfo {author}
  {\bibfnamefont{G.}~\bibnamefont{Rupper}}, \bibinfo {author}
  {\bibfnamefont{C.}~\bibnamefont{Ell}}, \bibinfo {author}
  {\bibfnamefont{O.~B.}\ \bibnamefont{Shchekin}},\ and\ \bibinfo {author}
  {\bibfnamefont{D.~G.}\ \bibnamefont{Deppe}},\ }%
  \bibfield{journal}{%
  \bibinfo {journal} {Nature}\ }%
  \textbf{\bibinfo {volume} {432}},\ \bibinfo {pages} {200} (\bibinfo {year}
  {2004})%
  \bibAnnoteFile{NoStop}{yoshie04a}%
\bibitem{hennessy07a}%
  \BibitemOpen
  \bibfield{author}{%
  \bibinfo {author} {\bibfnamefont{K.}~\bibnamefont{Hennessy}}, \bibinfo
  {author} {\bibfnamefont{A.}~\bibnamefont{Badolato}}, \bibinfo {author}
  {\bibfnamefont{M.}~\bibnamefont{Winger}}, \bibinfo {author}
  {\bibfnamefont{D.}~\bibnamefont{Gerace}}, \bibinfo {author}
  {\bibfnamefont{M.}~\bibnamefont{Atature}}, \bibinfo {author}
  {\bibfnamefont{S.}~\bibnamefont{Gulde}}, \bibinfo {author}
  {\bibfnamefont{S.}~\bibnamefont{{F\u alt}}}, \bibinfo {author}
  {\bibfnamefont{E.~L.}\ \bibnamefont{Hu}},\ and\ \bibinfo {author}
  {\bibfnamefont{A.}~\bibnamefont{{\u Imamo\=glu}}},\ }%
  \bibfield{journal}{%
  \bibinfo {journal} {Nature}\ }%
  \textbf{\bibinfo {volume} {445}},\ \bibinfo {pages} {896} (\bibinfo {year}
  {2007})%
  \bibAnnoteFile{NoStop}{hennessy07a}%
\bibitem{press07a}%
  \BibitemOpen
  \bibfield{author}{%
  \bibinfo {author} {\bibfnamefont{D.}~\bibnamefont{Press}}, \bibinfo {author}
  {\bibfnamefont{S.}~\bibnamefont{G\"otzinger}}, \bibinfo {author}
  {\bibfnamefont{S.}~\bibnamefont{Reitzenstein}}, \bibinfo {author}
  {\bibfnamefont{C.}~\bibnamefont{Hofmann}}, \bibinfo {author}
  {\bibfnamefont{A.}~\bibnamefont{L\"offler}}, \bibinfo {author}
  {\bibfnamefont{M.}~\bibnamefont{Kamp}}, \bibinfo {author}
  {\bibfnamefont{A.}~\bibnamefont{Forchel}},\ and\ \bibinfo {author}
  {\bibfnamefont{Y.}~\bibnamefont{Yamamoto}},\ }%
  \bibfield{journal}{%
  \bibinfo {journal} {Phys. Rev. Lett.}\ }%
  \textbf{\bibinfo {volume} {98}},\ \bibinfo {pages} {117402} (\bibinfo {year}
  {2007})%
  \bibAnnoteFile{NoStop}{press07a}%
\bibitem{nomura08a}%
  \BibitemOpen
  \bibfield{author}{%
  \bibinfo {author} {\bibfnamefont{M.}~\bibnamefont{Nomura}}, \bibinfo {author}
  {\bibfnamefont{Y.}~\bibnamefont{Ota}}, \bibinfo {author}
  {\bibfnamefont{N.}~\bibnamefont{Kumagai}}, \bibinfo {author}
  {\bibfnamefont{S.}~\bibnamefont{Iwamoto}},\ and\ \bibinfo {author}
  {\bibfnamefont{Y.}~\bibnamefont{Arakawa}},\ }%
  \bibfield{journal}{%
  \bibinfo {journal} {Appl. Phys. Express}\ }%
  \textbf{\bibinfo {volume} {1}},\ \bibinfo {pages} {072102} (\bibinfo {year}
  {2008})%
  \bibAnnoteFile{NoStop}{nomura08a}%
\bibitem{kistner08a}%
  \BibitemOpen
  \bibfield{author}{%
  \bibinfo {author} {\bibfnamefont{C.}~\bibnamefont{Kistner}}, \bibinfo
  {author} {\bibfnamefont{T.}~\bibnamefont{Heindel}}, \bibinfo {author}
  {\bibfnamefont{C.}~\bibnamefont{Schneider}}, \bibinfo {author}
  {\bibfnamefont{A.}~\bibnamefont{Rahimi-Iman}}, \bibinfo {author}
  {\bibfnamefont{S.}~\bibnamefont{Reitzenstein}}, \bibinfo {author}
  {\bibfnamefont{S.}~\bibnamefont{H\"ofling}},\ and\ \bibinfo {author}
  {\bibfnamefont{A.}~\bibnamefont{Forchel}},\ }%
  \bibfield{journal}{%
  \bibinfo {journal} {Opt. Express}\ }%
  \textbf{\bibinfo {volume} {16}},\ \bibinfo {pages} {15006} (\bibinfo {year}
  {2008})%
  \bibAnnoteFile{NoStop}{kistner08a}%
\bibitem{laucht09b}%
  \BibitemOpen
  \bibfield{author}{%
  \bibinfo {author} {\bibfnamefont{A.}~\bibnamefont{Laucht}}, \bibinfo {author}
  {\bibfnamefont{N.}~\bibnamefont{Hauke}}, \bibinfo {author}
  {\bibfnamefont{J.~M.}\ \bibnamefont{Villas-B\^oas}}, \bibinfo {author}
  {\bibfnamefont{F.}~\bibnamefont{Hofbauer}}, \bibinfo {author}
  {\bibfnamefont{G.}~\bibnamefont{B\"ohm}}, \bibinfo {author}
  {\bibfnamefont{M.}~\bibnamefont{Kaniber}},\ and\ \bibinfo {author}
  {\bibfnamefont{J.~J.}\ \bibnamefont{Finley}},\ }%
  \bibfield{journal}{%
  \bibinfo {journal} {Phys. Rev. Lett.}\ }%
  \textbf{\bibinfo {volume} {103}},\ \bibinfo {pages} {087405} (\bibinfo {year}
  {2009})%
  \bibAnnoteFile{NoStop}{laucht09b}%
\bibitem{dousse09a}%
  \BibitemOpen
  \bibfield{author}{%
  \bibinfo {author} {\bibfnamefont{A.}~\bibnamefont{Dousse}}, \bibinfo {author}
  {\bibfnamefont{J.}~\bibnamefont{Suffczy\'nski}}, \bibinfo {author}
  {\bibfnamefont{R.}~\bibnamefont{Braive}}, \bibinfo {author}
  {\bibfnamefont{A.}~\bibnamefont{Miard}}, \bibinfo {author}
  {\bibfnamefont{A.}~\bibnamefont{Lema\^itre}}, \bibinfo {author}
  {\bibfnamefont{I.}~\bibnamefont{Sagnes}}, \bibinfo {author}
  {\bibfnamefont{L.}~\bibnamefont{Lanco}}, \bibinfo {author}
  {\bibfnamefont{J.}~\bibnamefont{Bloch}}, \bibinfo {author}
  {\bibfnamefont{P.}~\bibnamefont{Voisin}},\ and\ \bibinfo {author}
  {\bibfnamefont{P.}~\bibnamefont{Senellart}},\ }%
  \bibfield{journal}{%
  \bibinfo {journal} {Appl. Phys. Lett.}\ }%
  \textbf{\bibinfo {volume} {94}},\ \bibinfo {pages} {121102} (\bibinfo {year}
  {2009})%
  \bibAnnoteFile{NoStop}{dousse09a}%
\bibitem{laussy08a}%
  \BibitemOpen
  \bibfield{author}{%
  \bibinfo {author} {\bibfnamefont{F.~P.}\ \bibnamefont{Laussy}}, \bibinfo
  {author} {\bibfnamefont{E.}~\bibnamefont{del Valle}},\ and\ \bibinfo {author}
  {\bibfnamefont{C.}~\bibnamefont{Tejedor}},\ }%
  \bibfield{journal}{%
  \bibinfo {journal} {Phys. Rev. Lett.}\ }%
  \textbf{\bibinfo {volume} {101}},\ \bibinfo {pages} {083601} (\bibinfo {year}
  {2008})%
  \bibAnnoteFile{NoStop}{laussy08a}%
\bibitem{laussy09a}%
  \BibitemOpen
  \bibfield{author}{%
  \bibinfo {author} {\bibfnamefont{F.~P.}\ \bibnamefont{Laussy}}, \bibinfo
  {author} {\bibfnamefont{E.}~\bibnamefont{del Valle}},\ and\ \bibinfo {author}
  {\bibfnamefont{C.}~\bibnamefont{Tejedor}},\ }%
  \bibfield{journal}{%
  \bibinfo {journal} {Phys. Rev. B}\ }%
  \textbf{\bibinfo {volume} {79}},\ \bibinfo {pages} {235325} (\bibinfo {year}
  {2009})%
  \bibAnnoteFile{NoStop}{laussy09a}%
\bibitem{averkiev09a}%
  \BibitemOpen
  \bibfield{author}{%
  \bibinfo {author} {\bibfnamefont{N.}~\bibnamefont{Averkiev}}, \bibinfo
  {author} {\bibfnamefont{M.}~\bibnamefont{Glazov}},\ and\ \bibinfo {author}
  {\bibfnamefont{A.}~\bibnamefont{Poddubny}},\ }%
  \bibfield{journal}{%
  \bibinfo {journal} {Sov. Phys. JETP}\ }%
  \textbf{\bibinfo {volume} {135}},\ \bibinfo {pages} {959} (\bibinfo {year}
  {2009})%
  \bibAnnoteFile{NoStop}{averkiev09a}%
\bibitem{carmichael89a}%
  \BibitemOpen
  \bibfield{author}{%
  \bibinfo {author} {\bibfnamefont{H.~J.}\ \bibnamefont{Carmichael}}, \bibinfo
  {author} {\bibfnamefont{R.~J.}\ \bibnamefont{Brecha}}, \bibinfo {author}
  {\bibfnamefont{M.~G.}\ \bibnamefont{Raizen}}, \bibinfo {author}
  {\bibfnamefont{H.~J.}\ \bibnamefont{Kimble}},\ and\ \bibinfo {author}
  {\bibfnamefont{P.~R.}\ \bibnamefont{Rice}},\ }%
  \bibfield{journal}{%
  \bibinfo {journal} {Phys. Rev. A}\ }%
  \textbf{\bibinfo {volume} {40}},\ \bibinfo {pages} {5516} (\bibinfo {year}
  {1989})%
  \bibAnnoteFile{NoStop}{carmichael89a}%
\bibitem{keldysh06a}%
  \BibitemOpen
  \bibfield{author}{%
  \bibinfo {author} {\bibfnamefont{L.~V.}\ \bibnamefont{Keldysh}}, \bibinfo
  {author} {\bibfnamefont{V.~D.}\ \bibnamefont{Kulakovskii}}, \bibinfo {author}
  {\bibfnamefont{S.}~\bibnamefont{Reitzenstein}}, \bibinfo {author}
  {\bibfnamefont{M.~N.}\ \bibnamefont{Makhonin}},\ and\ \bibinfo {author}
  {\bibfnamefont{A.}~\bibnamefont{Forchel}},\ }%
  \bibfield{journal}{%
  \bibinfo {journal} {Pis'ma ZhETF}\ }%
  \textbf{\bibinfo {volume} {84}},\ \bibinfo {pages} {584} (\bibinfo {year}
  {2006})%
  \bibAnnoteFile{NoStop}{keldysh06a}%
\bibitem{gonzaleztudela10a}%
  \BibitemOpen
  \bibfield{author}{%
  \bibinfo {author} {\bibfnamefont{A.}~\bibnamefont{Gonzalez-Tudela}}, \bibinfo
  {author} {\bibfnamefont{E.}~\bibnamefont{del Valle}}, \bibinfo {author}
  {\bibfnamefont{C.}~\bibnamefont{Tejedor}},\ and\ \bibinfo {author}
  {\bibfnamefont{F.}~\bibnamefont{Laussy}},\ }%
  \bibfield{journal}{%
  \bibinfo {journal} {Superlatt. Microstruct.}\ }%
  \textbf{\bibinfo {volume} {47}},\ \bibinfo {pages} {16} (\bibinfo {year}
  {2010})%
  \bibAnnoteFile{NoStop}{gonzaleztudela10a}%
\bibitem{laussy06b}%
  \BibitemOpen
  \bibfield{author}{%
  \bibinfo {author} {\bibfnamefont{F.~P.}\ \bibnamefont{Laussy}}, \bibinfo
  {author} {\bibfnamefont{M.~M.}\ \bibnamefont{Glazov}}, \bibinfo {author}
  {\bibfnamefont{A.}~\bibnamefont{Kavokin}}, \bibinfo {author}
  {\bibfnamefont{D.~M.}\ \bibnamefont{Whittaker}},\ and\ \bibinfo {author}
  {\bibfnamefont{G.}~\bibnamefont{Malpuech}},\ }%
  \bibfield{journal}{%
  \bibinfo {journal} {Phys. Rev. B}\ }%
  \textbf{\bibinfo {volume} {73}},\ \bibinfo {pages} {115343} (\bibinfo {year}
  {2006})%
  \bibAnnoteFile{NoStop}{laussy06b}%
\bibitem{laussy09b}%
  \BibitemOpen
  \bibfield{author}{%
  \bibinfo {author} {\bibfnamefont{F.~P.}\ \bibnamefont{Laussy}}\ and\ \bibinfo
  {author} {\bibfnamefont{E.}~\bibnamefont{del Valle}},\ }%
  \bibfield{journal}{%
  \bibinfo {journal} {AIP Conference Proceedings}\ }%
  \textbf{\bibinfo {volume} {1147}},\ \bibinfo {pages} {46} (\bibinfo {year}
  {2009})%
  \bibAnnoteFile{NoStop}{laussy09b}%
\bibitem{delvalle09a}%
  \BibitemOpen
  \bibfield{author}{%
  \bibinfo {author} {\bibfnamefont{E.}~\bibnamefont{del Valle}}, \bibinfo
  {author} {\bibfnamefont{F.~P.}\ \bibnamefont{Laussy}},\ and\ \bibinfo
  {author} {\bibfnamefont{C.}~\bibnamefont{Tejedor}},\ }%
  \bibfield{journal}{%
  \bibinfo {journal} {Phys. Rev. B}\ }%
  \textbf{\bibinfo {volume} {79}},\ \bibinfo {pages} {235326} (\bibinfo {year}
  {2009})%
  \bibAnnoteFile{NoStop}{delvalle09a}%
\bibitem{nomura10a}%
  \BibitemOpen
  \bibfield{author}{%
  \bibinfo {author} {\bibfnamefont{M.}~\bibnamefont{Nomura}}, \bibinfo {author}
  {\bibfnamefont{N.}~\bibnamefont{Kumagai}}, \bibinfo {author}
  {\bibfnamefont{S.}~\bibnamefont{Iwamoto}}, \bibinfo {author}
  {\bibfnamefont{Y.}~\bibnamefont{Ota}},\ and\ \bibinfo {author}
  {\bibfnamefont{Y.}~\bibnamefont{Arakawa}},\ }%
  \bibfield{journal}{%
  \bibinfo {journal} {Nat. Phys.}\ }%
  \textbf{\bibinfo {volume} {6}},\ \bibinfo {pages} {279} (\bibinfo {year}
  {2010})%
  \bibAnnoteFile{NoStop}{nomura10a}%
\bibitem{ota09b}%
  \BibitemOpen
  \bibfield{author}{%
  \bibinfo {author} {\bibfnamefont{Y.}~\bibnamefont{Ota}}, \bibinfo {author}
  {\bibfnamefont{N.}~\bibnamefont{Kumagai}}, \bibinfo {author}
  {\bibfnamefont{S.}~\bibnamefont{Ohkouchi}}, \bibinfo {author}
  {\bibfnamefont{M.}~\bibnamefont{Shirane}}, \bibinfo {author}
  {\bibfnamefont{M.}~\bibnamefont{Nomura}}, \bibinfo {author}
  {\bibfnamefont{S.}~\bibnamefont{Ishida}}, \bibinfo {author}
  {\bibfnamefont{S.}~\bibnamefont{Iwamoto}}, \bibinfo {author}
  {\bibfnamefont{S.}~\bibnamefont{Yorozu}},\ and\ \bibinfo {author}
  {\bibfnamefont{Y.}~\bibnamefont{Arakawa}},\ }%
  \bibfield{journal}{%
  \bibinfo {journal} {Appl. Phys. Express}\ }%
  \textbf{\bibinfo {volume} {2}},\ \bibinfo {pages} {122301} (\bibinfo {year}
  {2009})%
  \bibAnnoteFile{NoStop}{ota09b}%
\bibitem{khitrova06a}%
  \BibitemOpen
  \bibfield{author}{%
  \bibinfo {author} {\bibfnamefont{G.}~\bibnamefont{Khitrova}}, \bibinfo
  {author} {\bibfnamefont{H.~M.}\ \bibnamefont{Gibbs}}, \bibinfo {author}
  {\bibfnamefont{M.}~\bibnamefont{Kira}}, \bibinfo {author}
  {\bibfnamefont{S.~W.}\ \bibnamefont{Koch}},\ and\ \bibinfo {author}
  {\bibfnamefont{A.}~\bibnamefont{Scherer}},\ }%
  \bibfield{journal}{%
  \bibinfo {journal} {Nat. Phys.}\ }%
  \textbf{\bibinfo {volume} {2}},\ \bibinfo {pages} {81} (\bibinfo {year}
  {2006})%
  \bibAnnoteFile{NoStop}{khitrova06a}%
\bibitem{gonzaleztudela10b}%
  \BibitemOpen
  \bibfield{author}{%
  \bibinfo {author} {\bibfnamefont{A.}~\bibnamefont{Gonzalez-Tudela}}, \bibinfo
  {author} {\bibfnamefont{E.}~\bibnamefont{del Valle}}, \bibinfo {author}
  {\bibfnamefont{E.}~\bibnamefont{Cancellieri}}, \bibinfo {author}
  {\bibfnamefont{C.}~\bibnamefont{Tejedor}}, \bibinfo {author}
  {\bibfnamefont{D.}~\bibnamefont{Sanvitto}},\ and\ \bibinfo {author}
  {\bibfnamefont{F.~P.}\ \bibnamefont{Laussy}},\ }%
  \bibfield{journal}{%
  \bibinfo {journal} {Opt. Express}\ }%
  \textbf{\bibinfo {volume} {18}},\ \bibinfo {pages} {7002} (\bibinfo {year}
  {2010})%
  \bibAnnoteFile{NoStop}{gonzaleztudela10b}%
\end{thebibliography}%

\end{document}